\documentclass[pra,a4paper,preprintnumbers,aps,superscriptaddress,twocolumn,floatfix,showpacs,showkeys]{emulateapj}
\usepackage{graphicx}
\usepackage{float}
\usepackage{rotating}
\usepackage{epsfig,floatflt}
\newcommand{\dd}{{\rm d}}
\begin{document}
\title{An Improved Semi-Analytical Spherical Collapse Model for Non-linear Density Evolution}
\author{Douglas J. Shaw$^1$ and David F. Mota$^2$}
\affil{$^1$ DAMTP, Centre for Mathematical Sciences,
University of Cambridge,  Wilberforce Road, Cambridge CB3 0WA, UK}
\affil{$^2$ Institute for Theoretical Physics, University of Heidelberg, 69120 Heidelberg,Germany} 
%
\begin{abstract}
We derive a semi-analytical extension of the spherical collapse model of structure formation that takes account of the effects of deviations from spherical symmetry and shell crossing which are important in the non-linear regime.  Our model is designed so that it predicts a relation between the peculiar velocity and density contrast that agrees with the results of N-body simulations in the region where such a comparison can sensibly be made.  Prior to turnaround, when the unmodified spherical collapse model is expect to be a good approximation, the predictions of the two models coincide almost exactly.  The effects of a late time dominating dark energy component are also taken into account.  
The improved spherical collapse model is a useful tool when one requires a good approximation not just to the evolution of the density contrast but also its trajectory. Moreover, the analytical fitting formulae presented is  simple enough to be used anywhere where the standard spherical collapse might be used but with the advantage that it includes a realistic model of the effects of virialisation.
\end{abstract}

\maketitle
\section{Introduction}
The Spherical Collapse Model (hereafter SCM) developed by \citet{Gunn:1972} is perhaps the simplest model for the evolution of non-linear structure, and yet it has been shown to be remarkably successful when correctly interpreted.  However, despite the SCM's many success, it is ultimately flawed since it predicts that any overdensity of matter collapses to a singularity in a finite time.  Additionally, making the assumption of spherical symmetry, whilst simplifying, means abandoning the many interesting and important aspects of structure formation that result from deviations from spherical symmetry.  Indeed, these deviations play a crucial role in ultimately halting the collapse of the overdensity and the formation of virialized structures.  

The usual approach to virialisation in the SCM is to put it in \emph{by hand}; the collapse is simply halted once the virial radius has been reached.  This procedure, within an Einstein-de Sitter Universe, leads to the result that bound structures are formed when the non-linear overdensity is about 178 or, equivalently, the linear overdensity is approximately 1.68 \footnote{In dark energy scenarios these values change slightly depending on the model \citep{per,new,bag,nev}}.  Despite the fact that these figures are not too far away from what is actually observed in N-body simulations, the ad hoc nature of this approach means the SCM cannot be used to predict the precise manner in which the overdensity evolves; moreover, the SCM's prediction for the peculiar velocity becomes   virtually useless shortly after turnaround.

\citet{Engineer:2000} proposed a different and better motivated way in which the SCM could be extended to include virialisation.  Their idea was to alter the standard SCM evolution equation for the mean density contrast, $\bar{\delta}$, in such a way that stable structures would form.  Their modified evolution equation was constructed by adding a Taylor series in $1/\bar{\delta}$ to the standard equation: the idea being that these additional terms would encode all the effects due to shell-crossing and deviations from  spherical symmetry that occur for large $\bar{\delta}$.  The coefficients of the terms in Taylor series were chosen so as to provide a good approximation to the statistical density-contrast found from N-body simulations.  Whilst their improved SCM was found to agree fairly well with data from N-body simulations for $\bar{\delta} \gtrsim 15$, their solution for $\bar{\delta}$ is inaccurate in the linear regime, where spherical symmetry and hence also the unmodified SCM are expected to be good approximations.  

In this article we take a similar approach to \citet{Engineer:2000} and add terms to the standard SCM evolution equation so that for large $\bar{\delta}$ the evolution of the density contrast and the peculiar velocity agrees with the data from N-body simulations.  Importantly though, the analytical solutions, which we find, remain valid as $\bar{\delta} \rightarrow 0$ and agree almost exactly with the unmodified SCM prior to turnaround.  

\section{Modelling Non-Linear Structures}
We consider a spherical overdensity embedded in a background universe described by a  Friedman-Robertson-Walker (FRW) metric. In a spherically-symmetric system and in the absence of shell-crossing, the mass, $M$, inside each comoving spherical shell, with physical radius $R(t)$, remains constant. The mean density contrast inside shell therefore scales as:
\begin{equation}
1+\bar{\delta} = \lambda\frac{a^3(t)}{R(t)^3}, \label{deltaeqn}
\end{equation}
where $a(t)$ is the scale factor of the background FRW universe, and $\lambda$ is constant on each shell. Eq. (\ref{deltaeqn}) can also be used to define $R(t)$ for shells of constant $M$ when deviations from spherical symmetry occur. In these cases, however, the physical meaning of $R(t)$ is less clear. Following \citet{Engineer:2000} we continue to think of $R(t)$, as defined by Eq. (\ref{deltaeqn}), as the effective ``radius'' of each shell of constant $M$.

The peculiar velocity, $h_{SC}$, is defined by:
\begin{equation}
h_{SC} \equiv \frac{1}{3}\frac{\dd\ln(1+\bar{\delta})}{\dd \ln a} = \left(1 - \frac{\dot{R}}{H R}\right), \label{hsceqn}
\end{equation}
where $H = \dot{a}/a$ is the Hubble parameter.  In a matter dominated Universe the unmodified SCM predicts $h_{SC} = h_{SC}(\bar{\delta})$. Our key assumption in what follows is that this remains the case even when deviations from spherical symmetry are taken into account.  This assumption is supported by numerical studies of structure formation such as that conducted by \citet{Hamilton:1991} which we discuss below.  

For the moment we take the background universe to be matter dominated, however in Section \ref{degen} we generalize our results to  non-linear overdensities in more realistic universes which have recently transitioned to an epoch of dark energy domination. We define a new coordinate $\eta(\bar{\delta})$ by:
\begin{equation}
1+\bar{\delta} = \frac{9}{2}\frac{T(\eta)^2}{(1-\cos\eta)^3}, \label{deltaeqnT}
\end{equation}
for some function $T(\eta)$.  Now $M$ is constant on each shell and so, taking $t=t_{i}$, $R=R_{i}$ and $\bar{\delta}=\delta_{i}$ at some instance, we have 
$$M = \frac{2R_{i}^3(1+\delta_{i})}{9t_{i}^2} = \frac{2 R^3 (1+\bar{\delta})}{9 t^2}.$$
By considering this relation together with Eq. (\ref{deltaeqnT}) we see that:
\begin{eqnarray*}
R &=& \frac{R_{i}(1+\delta_{i})}{2\delta_{i} j(\eta, R_i)}\left(1-\cos(\eta)\right) \\
t &=& \frac{3 t_{i} (1+\delta_{i}) }{4 \delta_{i}^{3/2} j^{3/2}(\eta, R_i)}T(\eta).
\end{eqnarray*}
where $j(\eta, \delta_i)$ is some function of $\eta$  and $\delta_i$.  If $h_{SC} = h_{SC}(\bar{\delta})$ then $j(\eta, R_i)=j(\eta)$, and without loss of generality we can set $j(\eta) = 1$.  This gives:
\begin{equation}
h_{SC} = 1-\sqrt{\frac{1+\bar{\delta}}{2}} \frac{\sin \eta}{(1-\cos \eta)^{1/2}} \frac{\dd \tau}{\dd T}, \label{hsceqnT}
\end{equation}
where $\tau(\eta) = \eta - \sin \eta$, and:
\begin{eqnarray}
R &=& R(\eta) = \frac{R_{i}(1+\delta_i)}{2\delta_i}(1-\cos \eta), \label{Reta} \\ 
t &=& t(\eta) = \frac{3t_i(1+\delta_i)}{4\delta_i^{3/2}}T(\eta), \label{Teta}.
\end{eqnarray}
We reduce to the standard SCM in the limit where $T = \tau(\eta) =\eta - \sin \eta$ \citep{Pad:2002}, and it is clear that $h_{SC} = h_{SC}(\bar{\delta})$ in the unmodified SCM.  It is important to stress that if $h_{SC} = h_{SC}(\bar{\delta})$ then Eqs. (\ref{Reta}) and (\ref{Teta}) are the most general solutions for $R$ and $t$.

The radial acceleration of each shell is found to be:
\begin{equation}
\frac{\dd^2R}{\dd t^2} = - \frac{GM}{R^2}\left( T^{\prime\,-2}+ \sin \eta(1-\cos \eta) \frac{T^{\prime \prime}}{T^{\prime \, 3}}\right), \label{modeqn}
\end{equation}
where $T^{\prime} = \dd T / \dd \tau$, and $T^{\prime \prime} = \dd^2 T / \dd\tau^2$.  As should be expected, when $T^{\prime} = 1$ we reduce to the standard SCM equation for $R_{,tt}$.  

When deviations from spherical symmetry and the leading order effects of a gradient in the velocity dispersion are taken into account, \citet{Engineer:2000} showed that one should have:
\begin{eqnarray}
\frac{\dd^2 R}{\dd t^2} &=& -\frac{GM}{R^2} - \frac{H^2 R}{3} S \label{fluideqn} \\&=& -\frac{GM}{R^2}\left(1+ \frac{2}{3(1+\bar{\delta})}S\right), \nonumber
\end{eqnarray}
where $S(a, x) = a^2(\sigma^2 - 2\Omega^2) + f(a,x)$; $\sigma^2$ and $\Omega^2$ respectively quantify the shear and rotation of the fluid; $f(a,x)$ contains the lowest order contribution from velocity dispersion terms. Importantly, no matter what form $S(a,x)$ takes, if, as we have assumed $h_{SC}= h_{SC}(\bar{\delta})$, then we must have $S = S(\bar{\delta})$ and by comparing Eqs. (\ref{modeqn}) and (\ref{fluideqn}) we can clearly see that:
$$
S(\bar{\delta}) = \frac{3(1+\bar{\delta})}{2}\left(T^{\prime\,-2}-1+ \sin \eta (1-\cos \eta) \frac{T^{\prime \prime}}{T^{\prime \, 3}}\right).
$$
In principle, the form of both $S(\bar{\delta})$ and hence $T(\eta)$ can be found using the results of N-body simulations.  Unfortunately, however, making the required comparison with simulations is not as straightforward as one might expect it to be.  This is because the results of such simulations are given in terms of the statistical properties of the matter distribution rather than in terms of the mean density contrast, $\bar{\delta}$, and peculiar velocity, $h_{SC}$.  The statistical properties in question are the averaged two point correlation function, $\bar{\xi}$, and the averaged pair velocity, $h(a,x)$, which are given as defined by:
\begin{equation}
\bar{\xi} = \frac{3}{r^3}\int^{r}_{0} \xi(x,a) x^2\, dx; \qquad h(a,x) = -\frac{\langle v(a,x) \rangle}{\dot{a} x},
\end{equation}
where $\xi$ is the two-point correlation function and is defined to be the Fourier transform of the power spectrum, $P(k)$.  The assumption that $h(a,x)$ depends on $a$ and $x$ only through $\bar{\xi}$ i.e. $h(a,x) = h(\bar{\xi}(a,x))$ is common  in the literature (see \cite{Engineer:2000}, \cite{Moe:1995}) and it appears to have been confirmed by numerical simulations (see \cite{Hamilton:1991}, \cite{Peacock:1996}).  

The results of, for example, \citet{Hamilton:1991} can be used to construct the fitting formula for $h(\bar{\xi})$ \citep{Engineer:2000}. However, before we can make use of such a formula, we must relate the statistical quantities $\bar{\xi}$ and $h(\bar{\xi})$ to $\bar{\delta}$ and $h_{SC}(\bar{\delta})$.  It is well known that (\cite{Pad:1996},\cite{Pad:2002},\cite{Peebles:1980},\cite{Padeng},\cite{pop},\cite{pad}), on scales smaller than the size of the collapsing objects and around high density peaks:
\begin{equation}
\rho \simeq \rho_b(1+\xi).
\end{equation}
It follows that, in the non-linear regime, we have $\bar{\delta} \approx \bar{\xi}$.  This relationship between $\bar{\delta}$ and $\bar{\xi}$ was also used by \citet{Engineer:2000}, although it was, as it is here, the weakest part of the whole analysis. We only require that $\bar{\delta} \approx \bar{\xi}$ hold where it is expected to be a good approximation i.e. $\bar{\delta} \gtrsim 15$. As $\bar{\delta} \rightarrow \infty$ we assume that $\bar{\delta} \sim \bar{\xi}$. \citet{Peebles:1980} showed that $h(\bar{\xi})$ satisfies:
\begin{equation}
h = \frac{1}{3}\frac{1}{1+\bar{\xi}} \frac{\dd \bar{\xi}}{\dd \ln a}, \label{heqn}
\end{equation}
thus if $\bar{\xi} \approx \bar{\delta}$, then by comparing Eqs. (\ref{hsceqn}) and (\ref{heqn}), we see that $h \approx h_{SC}$.
It must be stressed that this second relation between $h_{SC}$ and $h$ is only valid if $\bar{\xi} \approx \bar{\delta}$, however whenever it does hold it implies that $h_{SC}$, is given by a function of $\bar{\delta}$ alone, i.e. $h_{SC}=h_{SC}(\bar{\delta})$.

The assumption that $\bar{\xi} \approx \bar{\delta}$ breaks down for small $\bar{\delta}$. Fortunately, when $\bar{\delta}$ is small, the unmodified SCM provides an accurate model. For times $t \gg t_{i}$, the unmodified SCM predicts that $h_{SC}=h_{SC}(\bar{\delta})$. Our key assumption that $h_{SC} = h_{SC}(\bar{\delta})$ is therefore expected to hold for almost all $\bar{\delta}$.  The assumption does break down for $t \sim t_i$, however this is entirely due to the decaying mode in $\bar{\delta}$ which is negligible for $t \gg t_i$.

\section{Constructing an improved SCM}
If $h_{SC} = h_{SC}(\bar{\delta})$ then all properties of a modified SCM are encoded in a single function $T(\eta)$.  Our aim is to combine the unmodified SCM and data from N-body simulations to find a fitting formula  for $T(\eta)$ that results in accurate predictions for $h_{SC}(\bar{\delta})$ in all regimes. 

Prior to turnaround we expect the standard SCM to be accurate and hence $T(\eta) \approx \eta - \sin \eta$. Furthermore, for the unmodified SCM to be accurate, at leading order, in the linear regime we must have $T(\eta) \sim \eta - \sin \eta + o(\eta^5)$ for small $\eta$.  When $\bar{\delta} \gtrsim 15$, we expect $h_{SC}(\bar{\delta}) \approx h(\bar{\xi})$ and $\bar{\xi} \approx \bar{\delta}$, and we may use $h(\xi)$ to extract the large $\bar{\delta}$ form of $T(\eta)$.  We describe how this is done below.  In a matter dominated universe, the linearly extrapolated mean two-point correlation function, $\bar{\xi}_{lin}$, scales as $\bar{\xi}_{lin} \propto a^2 \propto T^{4/3}$, therefore as $\bar{\delta} \rightarrow \infty$:
\begin{equation}
(1+\bar{\xi}) \approx (1+\bar{\delta}) = \frac{9}{2}\frac{T(\eta)^2}{(1-\cos(\eta))^3}, \label{Tetaeqn}
\end{equation}
and so
\begin{equation}
(1-\cos \eta) \approx (1-\cos \eta_{sim}) \equiv \frac{A \bar{\xi}_{lin}^{1/2}}{(1+\bar{\xi})^{1/3}}, \label{cosetaeqn}
\end{equation}
where $A$ is a constant and we treat it as a parameter to be fitted.  \citet{Hamilton:1991} found the following fitting formula for $\bar{\xi}_{lin}(\bar{\xi})$:
\begin{equation}
\bar{\xi}_{lin} = \bar{\xi}\left(\frac{1+ 0.0158\bar{\xi}^2 + 0.000115\bar{\xi}^3}{1+0.926\bar{\xi}^2 -0.0743\bar{\xi}^3 + 0.0156\bar{\xi}^4}\right)^{1/3}.
\end{equation}
We define $\eta_{\infty} = \lim_{\bar{\delta} \rightarrow \infty}\eta$. By taking $\bar{\xi} \rightarrow \infty$ we find that:
\begin{equation}
A = 2.2668 (1- \cos \eta_{\infty}) = 2.2668 \left(\frac{2 R_{vir}}{R_{ta}}\right),
\end{equation}
where we have used $\cos \eta_{\infty} = 1- 2R_{vir}/R_{ta}$ which follows from Eq.(\ref{Reta}); $R_{vir}$ is the radius of the shell at virialisation, and $R_{ta}$ is its radius at turnaround. If spherical symmetry is assumed then the virial theorem in an Einstein de Sitter Universe gives $R_{vir} = R_{ta}/2$. This  relation is generally used when virialisation is placed by hand into the SCM. For comparsion, \citet{Hamilton:1991} found that $R_{ta}/R_{vir} \approx 1.8$ from their simulations.  We treat $R_{vir}/R_{ta}$ as a fitting parameter. Eq. (\ref{cosetaeqn}) provides $\eta = \eta(\bar{\xi} \approx \bar{\delta})$, and $T(\eta)$ may now be found using Eq. (\ref{Tetaeqn}):
\begin{equation}
T(\eta(\bar{\xi})) \approx T_{sim} \equiv \left(\frac{2}{9}\right)^{1/2}A^{3/2} \bar{\xi}_{lin}^{3/4}(\bar{\xi}), \label{Teqn2}
\end{equation}
with $\tau \approx \tau_{sim} = \eta_{sim} - \sin \eta_{sim}$; $\eta_{sim}$ is defined by Eq.(\ref{cosetaeqn}).  

We fit for the parameter $A$ (or equivalently $R_{ta}/R_{vir}$) by considering some important physical constraints on the behaviour of $T(\eta)$. Since the effect of deviations from spherical symmetry is to slow down the collapse of the overdensity, it follows from Eq.(\ref{modeqn}) that:
\begin{equation}
T^{\prime\, 2} - \sin \eta (1-\cos \eta) \frac{T^{\prime \prime}}{T^{\prime}} \geq 1. \label{ineq1}
\end{equation}
Furthermore, the unaltered SCM should be a good approximation up to around turnaround (when $\eta = \pi$) and so we must have that $T^{\prime} \approx 1$, $T^{\prime \prime} \ll T^{\prime}$ for $\eta \lesssim \pi$.  As $\tau \rightarrow \eta_{\infty} -\sin\eta_{\infty}$, we must have $T \rightarrow \infty$ and so $T^{\prime} > 0$ as $\eta \rightarrow \eta_{\infty}$. Eq. (\ref{ineq1}) implies that, for $\eta > \pi$, we cannot have $0 < T^{\prime} < 1$ and $T^{\prime \prime} < 0$, and so it follows that $T^{\prime} \geq 1$ for $\eta > \pi$. It follows that $T \geq \tau$ everywhere.  This condition implies that we should choose $R_{vir}/R_{ta}$, and hence $A$, so that $T_{sim}$ is always greater than $\tau_{sim} \equiv \eta_{sim}-\sin\eta_{sim}$ for $\bar{\xi} \gtrsim 15$, which roughly corresponds to $\eta_{sim} \gtrsim 3.8$, $\tau_{sim} \gtrsim 4.4$. Moreover, since we want $T \rightarrow \tau$ as $\eta \rightarrow 0$, we choose $R_{vir}/R_{ta}$ so that, at the minimum of $T_{sim}$, $T_{sim} = \tau_{sim}$.  This requirement gives:
$$ 
\frac{R_{vir}}{R_{ta}} = 0.5896.
$$
We now use $T_{sim}$ to find a fitting formula for $T(\eta)$ that agrees with the unmodified SCM at early times (i.e. $T \sim \eta - \sin \eta$ for small $\eta$).
\section{Results}
We find that the fitting formula:
\begin{equation}
T(\eta(\tau)) = \tau + \frac{3.468(\tau_f - \tau)^{-1/2}\exp\left(-\frac{15(\tau_f - \tau)}{\tau}\right)}{(1+0.8(\tau_f-\tau)^{1/2}-0.4(\tau_f - \tau))},\label{Tfit}
\end{equation}
where $\tau_{f} = 5.516$, provides an excellent fit to the form of $T(\eta(\tau))$ derived from the simulations of \citet{Hamilton:1991}, i.e. $T \approx T_{sim}$, in the range $\delta \gtrsim 15$. It also provides an evolution of $\bar{\delta}$ that matches up smoothly to the one predicted by the standard SCM in the region where we expect it to provide an accurate approximation, i.e. prior to turnaround $\delta \lesssim 5$.  
\begin{figure}[ht]
\begin{center}
\includegraphics[angle=0,width=9.5cm]{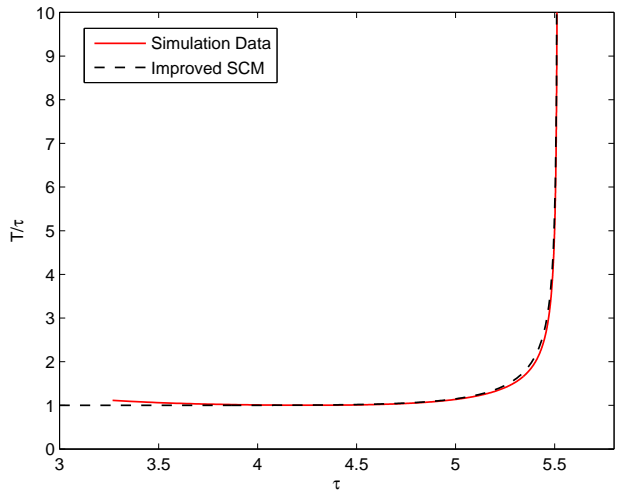} 
\caption{Plot of $T/\tau$ versus $\tau$.  The solid red line is $T/\tau=T_{sim}/\tau_{sim}$ against $\tau = \tau_{sim}$, whereas the dashed black line shows $T(\tau)/\tau$ in our improved SCM and uses the fitting formula for $T(\tau)$, Eq. (\ref{Tfit}), to evaluate $T/\tau$.  As required, we  see that in the region $\tau \gtrsim 4.4 \Rightarrow \bar{\delta} \gtrsim 15$, our improved SCM is a very good approximation to the simulation data.  For small values of $\tau$, we see that $T/\tau \rightarrow 1$, as is required, in our improved SCM.}
\label{fig1}
\end{center}
\end{figure}

In FIG \ref{fig1} we plot $T_{sim}$ against $\tau_{sim}$, and our fitting formula for $T(\eta(\tau))$ against $\tau$.  We see that, in the range $\bar{\delta}\gtrsim 15$, $\tau \gtrsim 4.4$ where we expect $\bar{\xi} \approx \bar{\delta}$ and hence $T_{sim} \approx T$, the fit is indeed very good.  The parameters in the formula for $T(\eta(\tau))$ have been chosen so that as $\tau \rightarrow \tau_{f}$, the leading order terms in the asymptotic expansions of both $T(\eta(\tau))$ and $T_{sim}(\eta_{sim}(\tau))$ match.
\begin{figure}[ht]
\begin{center}
\includegraphics[angle=0,width=9.5cm]{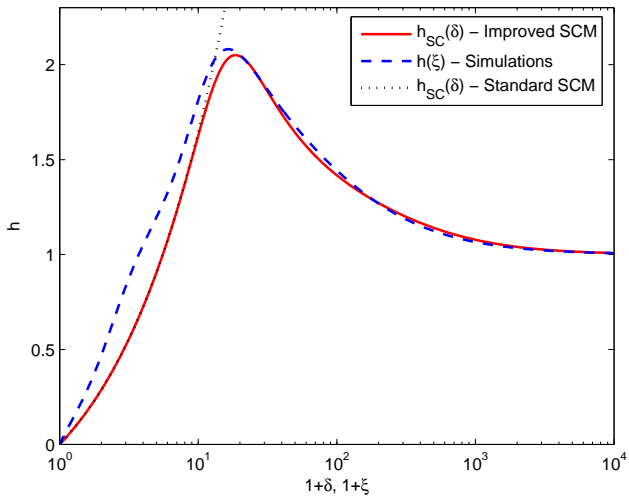}
 \caption{Relationship between $h$ and $\bar{\delta}$ in different models. The solid red line shows $h_{SC}(\bar{\delta})$ in our improved spherical collapse model.  The dotted black line is $h_{SC}(\bar{\delta})$ in the standard SCM, and the dashed  blue line shows $h(\bar{\xi})$ as seen in simulations \citep{Hamilton:1991}.  As required, our improved SCM gives an $h_{SC}$ that agrees with $h(\bar{\xi})$ very well in the region where we expect $\bar{\delta} \approx \bar{\xi}$ i.e. $\bar{\delta}\gtrsim 15$.  We also note that $h_{SC}(\bar{\delta})$, in the improved SCM, has the same asymptotic behaviour, to leading order, as $h(\bar{\xi}\approx \bar{\delta})$ in the limit $\bar{\delta} \rightarrow \infty$. Importantly, our improved SCM  model also agrees with the standard SCM in the region where it is expected to provide a very good approximation i.e. prior to the epoch of turnaround, $\bar{\delta} \lesssim 5.6$.}
 \label{fig2}
\end{center}
\end{figure}

In FIG \ref{fig2} we use Eqs. (\ref{deltaeqnT}) and (\ref{hsceqnT}) to plot $h_{SC}(\bar{\delta})$, and the fitting formula found by \citet{Hamilton:1991} to plot $h(\bar{\xi})$.  We also plot the unmodified SCM prediction for $h_{SC}(\bar{\delta})$.  It is clear from this plot that our fitting formula for $T$ gives an $h_{SC}(\bar{\delta})$ that is an excellent approximation to $h(\bar{\xi})$ (provided $\bar{\xi} \approx \bar{\delta}$) in the region $\bar{\delta} \gtrsim 20$.  As $\bar{\delta}, \bar{\xi} \rightarrow \infty$, the curves $h_{SC}(\bar{\delta})$ and $h(\bar{\xi})$ have the same leading order asymptotic behaviour. In the region $\bar{\delta} \gtrsim 15$, our model gives an $h_{SC}(\bar{\delta})$ that is always within $3\%$ of the fitting formula for $h(\bar{\xi})$ derived from simulations \citep{Hamilton:1991}.  Additionally, our fitting formula for $h_{SC}(\bar{\delta})$ agrees almost exactly with the predictions of the unmodified SCM prior to turnaround $\bar{\delta} < 5.6$.
\begin{figure}[ht]
\begin{center}
\includegraphics[angle=0,width=9.5cm]{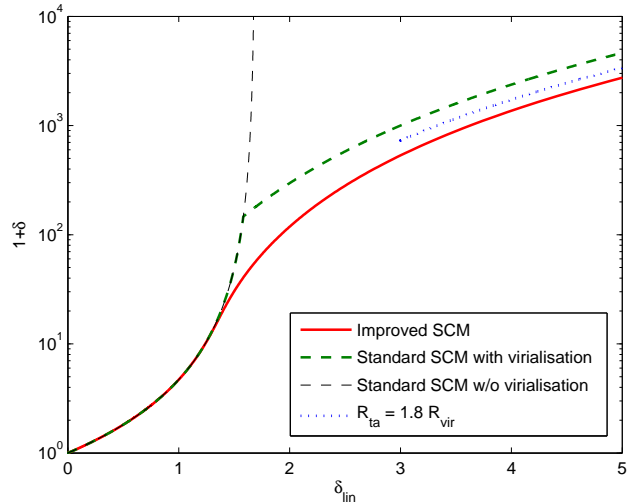}
 \caption{Plot of how the non-linear mean density contrast, $\bar{\delta}$, depends on the linear one, $\bar{\delta}_{lin}$.  The solid red line is the improved SCM developed here, the thick dashed green line  is the standard SCM with virialisation put in by hand and the thin dashed black line is the standard SCM without virialisation.  The dotted blue line shows the asymptotic behaviour of $\bar{\delta}(\bar{\delta}_{lin})$ that we would expect if $R_{ta}/R_{vir}=1.8$ as seen in the simulations of \citet{Hamilton:1991}. }
 \label{fig3}
\end{center}
\end{figure}
We found a best fit value of $R_{vir}/R_{ta} = 0.5896$, which is fairly close to the value of $0.5$ that is generally used in the spherical collapse model, despite the fact that we did not constrained it to be so.   For comparison, the improved spherical collapse developed by \citet{Engineer:2000} gave $R_{vir}/R_{ta} \approx 0.65$, and N-body simulations have been found to support $R_{vir}/R_{ta} \approx 0.56$ \citep{Hamilton:1991}.  Our value of $R_{vir}/R_{ta}$ is therefore a better approximation to the value found from simulations than that generally used in the unmodified SCM and that found in the model developed by \citet{Engineer:2000}.

In our improved SCM the linear density contrast is given by:
$$ 
\delta_{lin} = \frac{3}{5}\delta_{i} \left(\frac{t}{t_i}\right)^{2/3} = \frac{3}{5}\left(\frac{3}{4}\right)^{2/3} T(\eta)^{2/3}.
$$ 
When $\delta_{lin} = 1.6865$, which corresponds to the instant of collapse in the unmodified SCM, we find $\bar{\delta} = 54.65$ rather than the SCM's value of $178$.  We find that $\bar{\delta} = 200$ corresponds to $\delta_{lin} = 2.286$.  We compare the form of $\bar{\delta}(\bar{\delta}_{lin})$ found in our model to that predicted by the unmodified SCM model, with virialisation put in by hand, in FIG. \ref{fig3}.  We also show the divergent behaviour of $\bar{\delta}(\bar{\delta}_{lin})$ in the standard SCM without virialisation, and the asymptotic behaviour of $\bar{\delta}(\bar{\delta}_{lin})$ if $R_{ta} = 1.8 R_{vir}$ as suggested by simulations \citep{Hamilton:1991}.  The two major advantages of our improved SCM over the unmodified version are clearly visible in this plot:
\begin{itemize}
\item At late times the improved SCM provides a better approximation to the behaviour seen in N-body simulations than the unmodified SCM does.
\item $\bar{\delta}(\bar{\delta}_{lin})$ is smooth in the improved SCM. This is \emph{not} the case in the unmodified SCM when virialisation is put in by hand. The improved SCM therefore provides not only a good approximation to the evolution of density contrast, $\bar{\delta}$, but also to that of $\dd\bar{\delta}/\dd t$ and hence to that of the peculiar velocity.
\end{itemize}

\section{Including Dark Energy}\label{degen}
We have constructed an improved, semi-analytical SCM whose prediction for the dependence of peculiar velocity on the mean density contrast concurs with that extrapolated from simulations in the non-linear regime.  However, as mentioned above, we have done this for a matter dominated universe. It is well known, however, that the universe today is not matter dominated and a significant fraction of its total content is believed to be in the form of dark energy.  Fortunately though, the effects of this dark energy on the background universe are generally believed to only have become non-negligible relatively recently $z \lesssim 1.8$.  This means the local evolution of overdensities with $\bar{\delta} \gtrsim 100$ today has been matter-dominated, to a very good approximation, right up until the epoch of matter-radiation equality. Therefore the evolution of the density, $\rho$, and radial velocity, $v$, of a shell of matter that is well into the non-linear regime today is, to a good approximation, the same in a matter dominated background as it is in a background where today $\Omega_{m} \approx 0.27$ and $\Omega_{de} \approx 0.73$ for which the equation of state parameter for the dark energy, $w$, satisfies the current astronomical bound of $w = -1 \pm 0.1$ for $z < 1$ \citep{wbounds}.  It is therefore a fairly straightforward task to generalize our formulae for $\bar{\delta}$ and $h_{SC}$ to included background universes where $\Omega_{m}\approx 0.27$ provided that these are only applied to overdensities that  are large today.   We then find that:
\begin{eqnarray}
&&1+\bar{\delta}(\Omega_m,t) = \frac{9f(a) T(\eta)^2}{2(1-\cos\eta )^3}, \label{deltaom} \\
&&h_{SC}(\Omega_m,t) = 1- \Omega_{m}^{0.5}\sqrt{\frac{1+\bar{\delta}}{2(1-\cos\eta)}} \sin \eta \frac{\dd\tau}{\dd T} \label{hSCom}
\end{eqnarray}
where $\tau = \eta - \sin\eta$ as before, $T(\eta)$ is still given by Eq. (\ref{Tfit}) and
$$
f(a) = \frac{4}{9t^2 \Omega_m H^2}.
$$
The equations for $R(\eta)$ and $t(\eta)$ remain unchanged. 

If dark energy has behaved similarly to a cosmological constant in the recent past $(z < 1.8)$ then:
$$
f(a) \approx \frac{1-\Omega_m}{\Omega_{m} (\sinh^{-1}(\sqrt{(1-\Omega_{m})/\Omega_{m}}))^2} \approx \Omega_{m}^{-0.4}.
$$ 
Eqs. (\ref{deltaom}) and (\ref{hSCom}) combined with the fitting formula Eq. (\ref{Tfit}) provide a very good approximation to the evolution of overdensities of matter in a realistic universe provided that $\bar{\delta} \gtrsim 100$ today.  For smaller values of $\bar{\delta}$ our results are only accurate for $\Omega_{m} \approx 1$. However for $\bar{\delta} \lesssim 15$ the effects of deviations from spherical symmetry, which have been our primary concern in this article, are small enough to be ignored and the results derived using the unmodified SCM can be used with confidence.   Notice also that such fitting formulae and assumptions are strictly speaking only applicable to dark energy models that are not coupled to matter. If such a matter coupling is allowed then the whole process of structure formation, both in the linear and non-linear regimes, may change (see e.g. \cite{brook,lahav,van,maor,nunes,cham,cham1,tomi1,tomi2}).  This said, in many dark energy models, the matter coupling is constrained by experiments to be very small (relative to the coupling between matter and gravity), and as a result any alterations to the process of structure formation are similarly small.

\section{Conclusions}
In this article we have extended the Spherical Collapse Model so that it takes account of the effects of deviations from spherical symmetry and shell crossing which are important in the non-linear regime.  The key assumptions that we used when constructing our model was that $h_{SC}= h_{SC}(\bar{\delta})$ and that for $\bar{\delta} \gtrsim 15$, $\bar{\delta} \approx \bar{\xi}$. The latter assumption is probably the weakest link in the whole analysis, although  both assumptions are found commonly in the literature (see \cite{Engineer:2000, Moe:1995, Peebles:1980}).  Our improved SCM predicts a form for $h_{SC}(\bar{\delta})$ that is consistent with the results of N-body simulations in the regime where a comparison can sensibly be made ($\bar{\delta} \gtrsim 15$) and with the unmodified SCM prior to turnaround.  Analytical formulae for $\bar{\delta}$ and $h_{SC}$ in the improved model have been presented, and they essentially differ from the comparable formulae in the unmodified SCM only by the replacement of $\tau = \eta - \sin \eta$ with $T(\eta)$, which is given by Eq. (\ref{Tfit}).  The improved SCM is therefore simple enough to be used anywhere where the unmodified SCM might be used but with the advantage that it includes a realistic model of the effects of virialisation.

\acknowledgments DJS acknowledges support from PPARC. DFM acknowledges support from the A. Humboldt Foundation and the Research Council of Norway through project number 159637/V30.

\end{document}